\documentclass[journal,comsoc]{IEEEtran}
\usepackage[T1]{fontenc}
\usepackage{times}
\usepackage{soul}
\usepackage{url}
\usepackage[hidelinks]{hyperref}
\usepackage[utf8]{inputenc}
\usepackage{graphicx}
\usepackage{amsmath}
\usepackage{amsthm}
\usepackage{booktabs}
\usepackage{algorithm}
\usepackage{algorithmic}
\usepackage{multirow}
\usepackage{threeparttable}
\usepackage{amsfonts}
\usepackage{float}
\usepackage{url}
\usepackage{bm}
\usepackage[table,xcdraw]{xcolor}
\usepackage{subfigure}
\usepackage{stfloats}
\usepackage[justification=centering, small]{caption}
\usepackage[numbers,sort&compress]{natbib}
\newcommand{\cmmnt}[1]{}
\usepackage{tabularray}
\usepackage{bbding}
\usepackage{amssymb}

\usepackage{pifont}
\newcommand{\cmark}{\ding{51}}
\newcommand{\xmark}{\ding{55}}

\ifCLASSINFOpdf
\else
\fi
\usepackage{amsmath}
\usepackage[cmintegrals]{newtxmath}
\begin{document}
%
\title{Decoupling and Interacting Multi-Task Learning Network for Joint Speech and Accent Recognition}
%
%
%


\author{Qijie Shao,
        Pengcheng Guo,
        Jinghao Yan,
        Pengfei Hu,
        Lei Xie,~\IEEEmembership{Senior Member,~IEEE}

\thanks{Corresponding author: Lei Xie.}
  
\thanks{Qijie Shao, Pengcheng Guo, and Lei Xie are with the Audio, Speech and Language Processing Group (ASLP), School of Computer Science, Northwestern Polytechnical University, Xi’an 710072, China. Email: qjshao@npu-aslp.org (Qijie Shao), pcguo@nwpu-aslp.org (Pengcheng Guo), lxie@nwpu.edu.cn (Lei Xie).}

\thanks{Jinghao Yan and Pengfei Hu are with Tencent Research, Beijing 100193, China. Email: jinghaoyan@tencent.com (Jinghao Yan), alanpfhu@tencent.com (Pengfei Hu).}
}

%

%

\markboth{Journal of \LaTeX\ Class Files,~Vol.~14, No.~8, August~2015}%
{Wang \MakeLowercase{\textit{et al.}}: Bare Demo of IEEEtran.cls for IEEE Communications Society Journals}
%



\maketitle

\begin{abstract}
Accents, as variations from standard pronunciation, pose significant challenges for speech recognition systems. 
Although joint automatic speech recognition (ASR) and accent recognition (AR) training has been proven effective in handling multi-accent scenarios, 
current multi-task ASR-AR approaches overlook the granularity differences between tasks. Fine-grained units capture pronunciation-related accent characteristics, while coarse-grained units are better for learning linguistic information.
Moreover, an explicit interaction of two tasks can also provide complementary information and improve the performance of each other, but it is rarely used by existing approaches. 
In this paper, we propose a novel Decoupling and Interacting Multi-task Network (DIMNet) for joint speech and accent recognition, which is comprised of a connectionist temporal classification (CTC) branch, an AR branch, an ASR branch, and a bottom feature encoder. Specifically, AR and ASR are first decoupled by separated branches and two-granular modeling units to learn task-specific representations. The AR branch is from our previously proposed linguistic-acoustic bimodal AR model and the ASR branch is an encoder-decoder based Conformer model. 
Then, for the task interaction, the CTC branch provides aligned text for the AR task, while accent embeddings extracted from our AR model are incorporated into the ASR branch's encoder and decoder. 
Finally, during ASR inference, a cross-granular rescoring method is introduced to fuse the complementary information from the CTC and attention decoder after the decoupling. 
Our experiments on English and Chinese datasets demonstrate the effectiveness of the proposed model, which achieves ${21.45\%}$/${28.53\%}$ AR accuracy relative improvement and ${32.33\%}$/${14.55\%}$ ASR error rate relative reduction over a published standard baseline, respectively.

\end{abstract}


\section{Introduction}
\label{sec:Introduction}
\IEEEPARstart{A}{ccents} refer to the variations in standard pronunciation that are influenced by factors such as the speaker's education level, region, or native language~\cite{lippi2012english, huang2004accent}. For instance, when English is spoken with a Mandarin accent, it is considered a foreign accent, whereas Cantonese-influenced Mandarin is categorized as a regional accent. Despite the significant progress made in end-to-end automatic speech recognition (E2E ASR) in recent years, accents remain a significant challenge to user equality in speech recognition, leading to a decline in the performance of ASR models trained on standard pronunciation data~\cite{qian2022layer, shi2021accented, yang2022adaptive, deng2023adaptable}. As a result, there has been a growing interest in multi-accent ASR.

The multi-task ASR-AR framework has become a widely-used solution for overcoming the challenges posed by multi-accent~\cite{jain2018improved,zhang2021e2e,toshniwal2018multilingual}.
This framework typically consists of a shared encoder and two branches dedicated to ASR and AR tasks, respectively.
The shared encoder is responsible for extracting acoustic features from the input speech for both branches, and backpropagation from these two branches enables the shared encoder to learn how to extract both linguistic and accent representations.
Although it has been shown effective in joint modeling ASR and AR~\cite{zhang2021e2e}, the granularity difference between tasks suggests that simultaneously extracting features via a shared encoder may not be an optimal strategy.
Additionally, the ASR and AR branches currently exhibit limited interaction, impeding the full utilization of information from the opposing branch~\cite{he2019interactive,pang2020multi}.

\textbf{Granularity difference between ASR and AR:}
E2E ASR is a linguistic-related task, in which coarse-grained units like byte-pair encodings (BPEs)~\cite{sennrich2016neural} or characters are often used for better representing linguistic information, such as spelling and words. 
In contrast, the AR task is pronunciation-related and requires capturing small acoustic variations like pitch, intonation, and stress, where fine-grained modeling units such as phonemes or syllables are more suitable~\cite{huang2021aispeech}. 
Therefore, incorporating two-granularity modeling units into the multi-task ASR-AR is more appropriate. 
Nevertheless, the sequence lengths of the two-granularity units are inconsistent, making simultaneous encoding by a single shared encoder challenging.

\textbf{Improving AR with ASR:} 
Earlier AR models directly extracted low-level features such as frequency and timbre~\cite{najafian2014unsupervised,hanani2020spoken,turan2020achieving}, which could lead to overfitting on speaker and channel characteristics~\cite{chowdhury2020does, deng2021improving}.
Recent studies have shown that leveraging linguistic information from ASR can effectively mitigate the overfitting issue in AR tasks~\cite{gao2021end,jain2018improved,shi2021accented,zhang2021e2e}. 
Initializing the AR encoder with a pre-trained ASR encoder~\cite{shi2021accented} or jointly training a multi-task ASR-AR network~\cite{zhang2021e2e,sun2018domain} are the commonly used approaches, both of which have demonstrated their effectiveness on various datasets.
Different from implicitly integrating linguistic information into AR models, in our previous study~\cite{shao2022linguistic}, we proposed a novel linguistic-acoustic similarity-based accent shift (LASAS) AR model.
To estimate the accent shift of an accented speech utterance, we first map the frame-level aligned text to multiple accent-associated anchor spaces and then leverage the similarities between the acoustic embedding and those anchors as an accent shift.
Compared with pure acoustic embedding, the learned accent shift takes full advantage of both linguistic and acoustic information, which can effectively improve AR performance. 

\textbf{Improving ASR with AR:} 
Likewise, knowing the accent embedded in the speech utterance is also beneficial to speech recognition. Plenty of studies have explored using complementary accent information to improve ASR performance on accent speech~\cite{jain2018improved,imaizumi2020dialect}.
In the multi-task ASR-AR framework, accent embeddings from the AR branch can be leveraged to enhance accent information in the ASR branch. However, the effectiveness of different types of accent embeddings on ASR performance can vary significantly. The hidden states of the DNN-based accent classifier~\cite{qian2022layer} provide rich and stable utterance-level accent information, while the AR posterior probabilities are more straightforward and interpretable. On the other hand, the accent shifts in the LASAS~\cite{shao2022linguistic} method capture accent variations and provide more detailed information.
Moreover, incorporating accent embeddings into either the encoder~\cite{qian2022layer} or decoder~\cite{imaizumi2020dialect} of the ASR model allows for the model to adapt to variations in pronunciation or linguistics, leading to varying effects on ASR performance. 
Thus, comprehensive studies to investigate the interpretability and ASR performance of each approach are essential.

The objective of this study is to overcome the challenge of \textit{unit-granularity differences} and promote \textit{full interaction} between the ASR and AR branches in a multi-task setting.
To achieve this, we present a Decoupling and Interacting Multi-task Network (DIMNet) for joint speech and accent recognition, which includes a connectionist temporal classification (CTC)~\cite{graves2006connectionist} branch, an AR branch, an ASR branch, and a bottom feature encoder. 
Specifically, AR and ASR are first decoupled by separated branches and two-granular modeling units to learn task-specific representations. The AR branch is from our previously proposed LASAS~\cite{shao2022linguistic} AR model and the ASR branch is an encoder-decoder-based Conformer~\cite{gulati2020conformer} model. Then, for the task interaction, the CTC branch is optimized with the same modeling units as the AR branch to provide linguistic features for the AR task, while latent accent embeddings extracted from our AR model are used to improve the ASR branch. We also conduct comprehensive studies to explore the choice of accent embeddings and the fusion strategies for the ASR branch. Finally, a cross-granular rescoring method is introduced to effectively fuse the probabilities from CTC and attention decoder during ASR inference.
Our experiments on English and Chinese datasets demonstrate the effectiveness of the proposed model, which achieves ${21.45\%}$/${28.53\%}$ AR accuracy relative improvement and ${32.33\%}$/${14.55\%}$ ASR error rate relative reduction over a published standard baseline, respectively.

\section{Related Works}
\label{sec:Related Works}
In this section, we present a brief summary of multi-accent ASR-AR frameworks and the applications of two-granularity modeling units.

\subsection{Multi-Accent ASR-AR}
Current approaches for multi-accent ASR-AR can be classified into three categories: cascade~\cite{deng2021improving,qian2022layer}, multi-task~\cite{zhang2021accent,zhang2021e2e,hu2021redat,viglino2019end,yang2018joint}, and single-task~\cite{gao2021end,yadavalli2022multi}.
For the cascade ASR-AR framework, ASR and AR models are trained separately and used in a sequential manner.
Typically, an AR model is first trained to extract accent features from the input speech, which is then utilized to assist the ASR model. 
Deng~\textit{et al.}~\cite{deng2021improving} proposed a cascade ASR-AR based on pre-trained wav2vec 2.0~\cite{baevski2020wav2vec}, achieving state-of-the-art (SOTA) performance on the AESRC dataset~\cite{shi2021accented} due to the powerful acoustic modeling capabilities of wav2vec 2.0. In~\cite{qian2022layer,gong2022layer}, Gong and Qian~\textit{et al.} also obtained competitive results using a cascade ASR-AR scheme.
Their AR component has a phonetic posteriorgrams (PPG) extractor and a time delay neural network (TDNN) based classifier~\cite{desplanques2020ecapa, waibel1989phoneme}, while the ASR component incorporates accent information into the encoder through adapter layers in a CTC/attention framework.
Although each model could be optimized with a large amount of in-domain data, the cascade strategy also introduces inevitable error propagation and increased computation complexity.
For the multi-task structure, a shared encoder is generally utilized to simultaneously extract accent and linguistic information. In~\cite{zhang2021accent}, Zhang~\textit{et al.} regarded ASR as an auxiliary task for AR.
By extracting phoneme-level accent variations, their method effectively improves AR performance, which provides evidence of ASR's helpfulness in AR tasks.
In~\cite{zhang2021e2e}, Zhang~\textit{et al.} incorporated an AR branch into a CTC/attention ASR and used a shared encoder to simultaneously learn accent and linguistic representations, leading to improved ASR adaptation to accents. 
Finally, single-task ASR-AR splices accent and text labels together, using a unified encoder-decoder structure to predict two kinds of labels simultaneously. Following this direction, Gao~\textit{et al.}~\cite{gao2021end} proposed a single-task scheme for accent prediction, which extends the output token list by inserting accent labels into the text transcripts and yields good results without modifying the E2E model structure.

In contrast to existing multi-task ASR-AR approaches, our proposed DIMNet incorporates decoupling and interacting, resulting in improved transfer and fusion of linguistic and acoustic information. To achieve this, we employ distinct branches and two granular modeling units to decouple the AR and ASR tasks, allowing them to focus on pronunciation-related and semantic-related aspects, respectively. Additionally, we introduce interacting between these branches through LASAS AR and encoder-decoder accent embedding fusion within DIMNet.

\subsection{Two-granularity Unit Modeling}
Research on the use of two-granularity modeling units in ASR-AR frameworks is relatively scarce. 
One notable approach was proposed by Rao~\textit{et al.}~\cite{rao2017multi}, which involved a multi-accent ASR that utilized phoneme-grapheme two-granularity modeling units. This model generated both phoneme and grapheme (a-z) outputs with multiple CTC decoders added to the encoder intermediate layers and a final CTC decoder stacked after the encoder. 
The number of CTCs in the intermediate layer was consistent with the accent types. However, unlike typical two-granularity schemes, the final outputs of this model were graphemes, which are finer-grained than the phoneme-based middle outputs. Two-granularity modeling units have also been used in pure ASR studies. For example, Chan~\textit{et al.}~\cite{chan2016online} introduced syllable/character units to an early attention-based ASR model with two decoders, but only the character decoder was used for inference. Other studies~\cite{zhou2018comparison,chen2018modular,zhou2018syllable,yuan2021decoupling,wang2021cascade} employed cascade audio-to-phoneme (A2P) and phoneme-to-word (P2W) schemes. Their experiments demonstrated that, compared with the acoustic embedding of the encoder, the pure-text phoneme inputs for P2W lack sufficient acoustic information. Thus, this two-stage independent structure leads to error accumulation, which requires a large amount of P2W data to mitigate. To address this problem, Zhang~\textit{et al.}~\cite{zhang2021decoupling} fused a P2W model and an attention decoder together in autoregressive decoding, which fully utilized the phoneme text and acoustic embedding. Alternatively, Yang~\textit{et al.}~\cite{yang2022multi} directly utilized a Transformer decoder~\cite{vaswani2017attention} to complete the P2W transcription. This method is elegant and concise but only supports two-granularity modeling units with the same sequence lengths, such as Chinese syllables and characters.

Unlike the aforementioned methods that directly apply fine-grained units to the ASR task, our DIMNet incorporates these units into the AR task, resulting in significant enhancements. Additionally, DIMNet avoids the inclusion of cascade A2P and P2W structure, effectively mitigating the accumulation of errors.

\subsection{Decoupling and Interacting Multi-Task Learning}
The efficacy of decoupling and interacting multi-task learning has been evident in various domains, including speech and speaker recognition~\cite{tang2016collaborative}, object detection~\cite{pang2020multi}, and sentiment analysis~\cite{he2019interactive}. 
Tang~\textit{et al.}~\cite{tang2016collaborative} introduced a collaborative joint training approach for speech and speaker recognition, where the output of one task is backpropagated to the other task, resulting in enhanced performance on both speech and speaker recognition tasks compared to single-task systems. 
Furthermore, similar decoupling and interacting methodologies have been employed to extract information at different scales.
Pang~\textit{et al.}~\cite{pang2020multi} proposed aggregate interaction modules to integrate features from adjacent levels, enabling the extraction of multi-scale image features. 
And He~\textit{et al.}~\cite{he2019interactive} presented an interactive multi-task learning network capable of jointly learning token-level and document-level sentiment information.

In the AR-ASR tasks, a major challenge in the interaction process is the limited ability of AR to directly leverage linguistic information from ASR. Our DIMNet tackles this challenge by incorporating the LASAS AR model, which utilizes ASR aligned text as one of its inputs. This distinguishes DIMNet from other approaches in decoupling and interacting multi-task learning methods.

\begin{figure*}[htp]
\centering
\vspace{0em}
\includegraphics[width=40em]{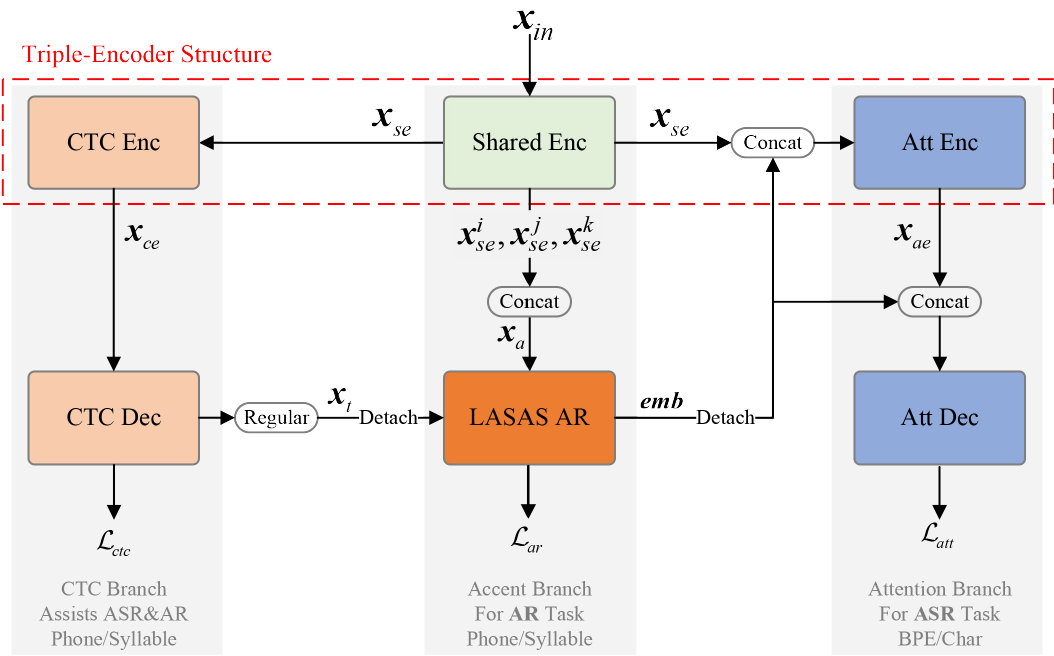}
\vspace{0em}
\caption{The framework overview of DIMNet.}
\vspace{-1em}
\label{fig:system_framework}
\end{figure*}

\section{Method}
\label{sec:Method}
Fig.~\ref{fig:system_framework} overviews the architecture of our proposed DIMNet, which uses a CTC/attention ASR~\cite{yao2021wenet} model as the backbone. In addition, we incorporate a LASAS AR~\cite{shao2022linguistic} model and a triple-encoder structure, resulting in three branches within the DIMNet. The accent and attention branches are dedicated to the AR and ASR tasks, respectively, while the CTC branch provides complementary information to assist the other two branches. We will illustrate these components of the DIMNet in the following subsections.

\subsection{Decoupling of AR and ASR}
Decoupling AR and ASR tasks can improve their respective performance by allowing for independent modeling of accent and linguistic information. Furthermore, a decoupled framework can enhance the clarity and interpretability of the interaction between the two tasks. 

For decoupling the DIMNet to allow each task to focus on different levels of information, we introduce a two-granularity modeling approach. Specifically, we use fine-grained units for the CTC branch to obtain aligned text as the inputs of the accent branch, and use coarse-grained units for the attention branch instead. 
The fine-grained units discussed here need to be pronunciation-related, such as the ARPAbet phoneme set~\cite{klautau2001arpabet} for English and Pinyin syllables for Chinese. On the contrary, for coarse-grained units, it is important for them to be semantic-related. In this paper, we use BPE and Char as coarse-grained units for English and Chinese, respectively.
The use of pronunciation-related fine-grained units allows the accent characteristics to be captured effectively in the AR task. On the other hand, semantic-related coarse-grained units are helpful in providing contextual linguistic information for the ASR task. Hence, the two-granularity decoupling helps enhance the performance of both tasks simultaneously.

As mentioned in Section~\ref{sec:Introduction}, sequence length inconsistency is a significant challenge in two-granularity modeling. To address this issue, we introduce the triple-encoder structure in the DIMNet. In this approach, a shared encoder is used to learn pronunciation-related shallow acoustic information, while two lightweight encoders are used for the CTC and attention branches to extract linguistic information in different granularity. The outputs of the triple encoders can be denoted as:
\begin{equation}
\label{eq:triple_encoder}
    \left\{\begin{matrix}
    \mathbf{x}_{se}=\text{Encoder}_{shared}(\mathbf{x}_{in}), \hfill \\
    \mathbf{x}_{ce}=\text{Encoder}_{ctc}(\mathbf{x}_{se}), \hfill \\
    \mathbf{x}_{ae}=\text{Encoder}_{att}(\text{Concat}(\mathbf{x}_{se},\ \mathbf{emb})), \hfill \\
    \end{matrix}\right.
\end{equation}
where ${\mathbf{x}_{in}}$ is the acoustic features of the input speech, which can be MFCC or Fbank, and ${\mathbf{x}_{se}}$ denotes the last layer's output of the shared encoder. The ${\mathbf{emb}}$ is an accent embedding extracted from the accent branch, which will be introduced in Section~\ref{sec:Accent Embedding Choosing and Transfer}.

By stacking additional encoders after the shared encoder, our approach can alleviate the sequence-length inconsistency of the two-granularity unit modeling. And then, the computation of the losses for the CTC and attention decoders are denoted as:
\begin{equation}
\label{eq:ctc_branch}
    \mathcal{L}_{ctc}=\text{CTC}(\mathbf{x}_{ce},\ \mathbf{y}_{f}), \hfill
\end{equation}
\begin{equation}
\label{eq:attention_branch}
    \left\{\begin{matrix} P(\mathbf{y}_{c}|\mathbf{x}_{in})=\text{Decoder}_{att}(\text{Concat}(\mathbf{x}_{ae},\ \mathbf{emb}),\ \mathbf{y}_{c}), \hfill \\
    \mathcal{L}_{att}=\text{CrossEntropy}(P(\mathbf{y}_{c}|\mathbf{x}_{in}),\ \mathbf{y}_{c}), \hfill
    \end{matrix}\right.
\end{equation}
where ${\mathbf{y}_{f}}$ and ${\mathbf{y}_{c}}$ are the transcription labels with fine-grained and coarse-grained units, respectively. The ${\mathbf{y}_{f}}$ could be translated to ${\mathbf{y}_{c}}$ with a lexicon. ${P(\mathbf{y}_{c}|\mathbf{x}_{in})}$ is the attention posterior probabilities of the given labels ${\mathbf{y}_{c}}$.

The total loss of our multi-task ASR-AR consists of the ASR loss ${\mathcal{L}_{att}}$, the CTC loss ${\mathcal{L}_{ctc}}$, and the AR loss ${\mathcal{L}_{ar}}$, which can be formulated as:
\begin{equation}
\label{eq:loss_ar_asr}
\mathcal{L}_{ASR-AR}=\mathcal{L}_{att}\ +\ \lambda_{1}\mathcal{L}_{ctc}\ +\ \lambda_{2}\mathcal{L}_{ar},
\end{equation}
where ${\lambda_{1}}$ and ${\lambda_{2}}$ are tunable hyperparameters. The details of ${\mathcal{L}_{ar}}$ will be introduced in Section~\ref{sec:LASAS Accent Recognition}.

\subsection{Improving AR with ASR}
\label{sec:LASAS Accent Recognition}
The traditional AR models only utilize acoustic features as input. When being integrated into a multi-task ASR-AR, their ability to fully interact with the ASR task is limited, which can hinder their performance. In our previous study~\cite{shao2022linguistic}, we proposed an AR model named LASAS that uses aligned text and acoustic features as input. This model explicitly and fully utilizes linguistic information, resulting in a significant improvement in AR performance. To enhance the ASR's assistance to AR, in this study, we first introduce LASAS into the multi-task framework by feeding it with the aligned text output from the CTC decoder and acoustic features output from the shared encoder. In addition, we further improve LASAS to make it adaptable to the multi-task ASR-AR.
\begin{figure}[htp]
\centering
\vspace{0em}
\includegraphics[width=24em]{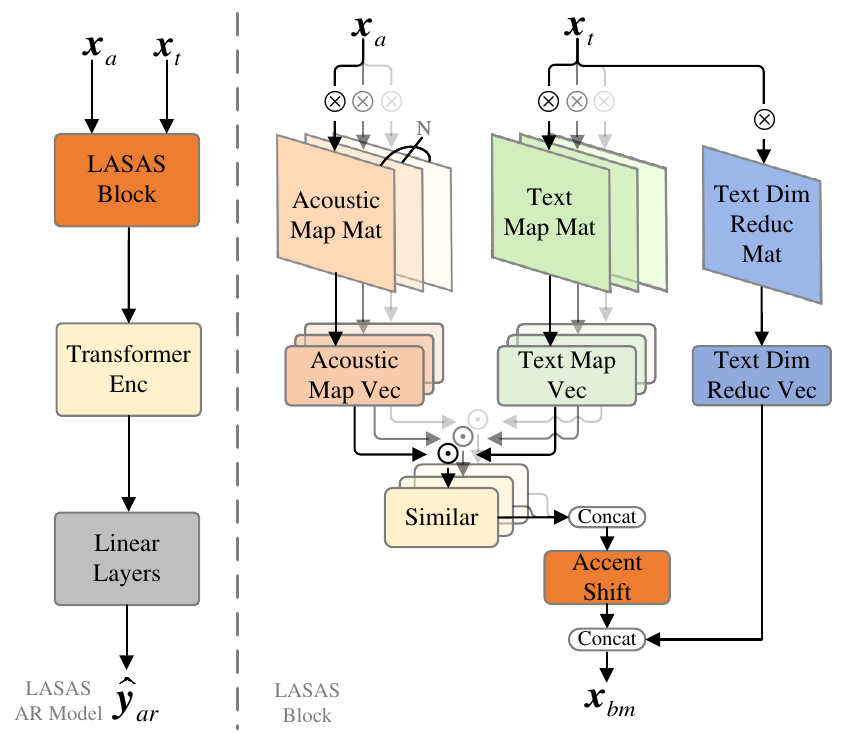}
\vspace{0em}
\caption{Details of LASAS block.}
\vspace{-1em}
\label{fig:lasas}
\end{figure}

The left part of Fig.~\ref{fig:lasas} shows the structure of the LASAS AR model, which consists of a LASAS block, a Transformer encoder~\cite{vaswani2017attention}, and several linear layers. The details of the LASAS block are depicted in the right part of Fig.~\ref{fig:lasas}, where the acoustic embedding ${\mathbf{x}_{a}}$ and frame-level aligned text ${\mathbf{x}_{t}}$ serve as inputs to LASAS, provided by the shared encoder and CTC decoder, respectively. The computation of  ${\mathbf{x}_{a}}$ and ${\mathbf{x}_{t}}$ can be denoted as:
\begin{equation}
    \left\{\begin{matrix}
    \mathbf{x}_{a}=\text{Concat}(\mathbf{x}_{se}^i,\ \mathbf{x}_{se}^j,\ \mathbf{x}_{se}^k)), \hfill \\
    \mathbf{x}_{t}=\text{Regular}(\text{GreedySearch}_{ctc}(\mathbf{x}_{ce})), \hfill
    \end{matrix}\right.
\end{equation}
where ${\mathbf{x}_{se}^i}$, ${\mathbf{x}_{se}^j}$, and ${\mathbf{x}_{se}^k}$ denote the outputs of the shared encoder's layers at the 1/3, 2/3, and 3/3 positions, respectively. For instance, if the shared encoder comprises 9 layers, then ${\mathbf{x}_{se}^i}$, ${\mathbf{x}_{se}^j}$, and ${\mathbf{x}_{se}^k}$ correspond to the outputs of the 3-rd, 6-th, and 9-th layers, respectively.
The greedy search used here retains both the blanks and repeated tokens of CTC, which differs from the conventional scheme. And ${\text{Regular}}$ means to replace blanks with subsequent predicted token IDs, constructing a frame-level aligned text with only repeated token IDs. 
It is worth noting that we use greedy search instead of prefix beam search to predict only one aligned text for the accent branch in both the training and inference stages. If prefix beam search is used during the training process, the time consumption will skyrocket to an unacceptable level.
Moreover, through experimental observations, we find that the AR performance of the model does not exhibit significant improvement (only a relative ${0.01\%}$) when using greedy search during training and switching to prefix beam search during inference. This lack of improvement can be attributed to the mismatch between the training process and the utilization of prefix beam search, despite its ability to generate text with a lower error rate.

The right part of Fig.~\ref{fig:lasas} illustrates the details of the LASAS block, which begins by mapping an aligned text vector ${\mathbf{x}_{t}}$ to multiple Euclidean spaces as anchors ${\mathbf{v}_{t}}$. These anchors are related to the speech content and aligned with the speech at a frame level. Next, we map an acoustic embedding ${\mathbf{x}_{a}}$ to the same dimension as the text anchor ${\mathbf{v}_{t}}$, denoted as ${\mathbf{v}_{a}}$. Using a scaled dot-product frame by frame, we obtain a similarity value ${s^{i}}$ between an anchor ${\mathbf{v}_{t}^i}$ and a mapped acoustic embedding ${\mathbf{v}_{a}^i}$. Finally, we concatenate a set of similarities to form ${\mathbf{s}}$, which reflects the shift directions and similarity degrees of different accents. The computation of the accent shift ${\mathbf{s}}$ can be denoted as:
\begin{equation}
    \left\{\begin{matrix}
    \mathbf{v}_{t}^i=\mathbf{x}_{t}\ \cdot\ \mathbf{W}_{t}^i, \hfill \\
    \mathbf{v}_{a}^i=\mathbf{x}_{a}\ \cdot\ \mathbf{W}_{a}^i, \hfill \\
    s^{i}=\text{DotProd}(\mathbf{v}_{a}^i,\ \mathbf{v}_{t}^i)\ /\ {\sqrt{d_{k}}}, \hfill \\
    \mathbf{s}=\text{Concat}(s^{1},\ s^{2},...,\ s^{N}), \hfill
    \end{matrix}\right.
\end{equation}
where ${i\in [1,N]}$ and ${N}$ is the number of mapping spaces. The ${\mathbf{W}_{t}^i\in\mathbb{R}^{D_{2}\times C}}$ and ${\mathbf{W}_{a}^i\in\mathbb{R}^{D_{1}\times C}}$ are mapping matrix. The $D_1$, $D_2$, and $C$ are the dimensions of features.

Since the accent shift ${\mathbf{s}}$ is a relative representation, a textual reference is necessary for the subsequent classifier. In order to represent pure textual information, we establish a dimension reduction matrix ${\mathbf{W}_{td}}$ to reduce the dimension of the input text OneHot vector. By concatenating the accent shift ${\mathbf{x}_{bm}}$ and the dimension-reduced text ${\mathbf{v}_{td}}$, we can obtain a linguistic-acoustic bimodal representation ${\mathbf{x}_{bm}}$, which can then be used as input for the subsequent classifier. Given the aligned text vector $\mathbf{x}_{t}$ and accent shift ${\mathbf{s}}$, the bimodal representation ${\mathbf{x}_{bm}}$ can be calculated as:
\begin{equation}
    \left\{\begin{matrix}
    \mathbf{v}_{td}=\mathbf{x}_{t}\ \cdot\ \mathbf{W}_{td}, \hfill \\
    \mathbf{x}_{bm}=\text{Concat}(\mathbf{s},\ \mathbf{v}_{td}), \hfill
    \end{matrix}\right.
\end{equation}
where ${\mathbf{W}_{td}\in\mathbb{R}^{D_{2}\times (C-N)}}$, ${\mathbf{x}_{bm}\in\mathbb{R}^{T\times C}}$, and $T$ is the time steps of frames.

Finally, the classifier is composed of a lightweight Transformer encoder and multiple linear layers. The encoder is used to extract context-sensitive accent information. 
After passing through the classifier, we obtain either a frame-level or utterance-level accent prediction. Utterance-level accent prediction tends to achieve better performance on AR tasks, while frame-level accent is more suitable as an embedding to assist ASR tasks. We will analyze and discuss the choice of the level in Section~\ref{sec:Experiments Results}.
Computation of the final accent prediction can be denoted as:
\begin{equation}
    \left\{\begin{matrix}
    P(\mathbf{y}_{ar}|\mathbf{x}_{in})=\text{Softmax}(\text{Linear}(\text{Transformer}(\mathbf{x}_{bm}))), \hfill \\
    \hat{\mathbf{y}}_{ar}=\arg\max(P(\mathbf{y}_{ar}|\mathbf{x}_{in})), \hfill
    \end{matrix}\right.
\end{equation}
where ${P(\mathbf{y}_{ar}|\mathbf{x}_{in})}$ represents the posterior probabilities for accent, and ${\hat{\mathbf{y}}_{ar}}$ refers to the predicted accent outcome from the accent branch.

To improve the adaptation of the original LASAS AR model~\cite{shao2022linguistic} to the multi-task ASR-AR, we detach the aligned text and accent embedding, as depicted in Figure~\ref{fig:system_framework}. This decouples the accent branch and allows for separate optimization of the CTC/attention and accent branches during back-propagation without mutual interference. The decoupling reduces learning difficulty and improves the performance of both the accent and attention branches. 
The loss of the accent branch is denoted as:
\begin{equation}
   \mathcal{L}_{ar}=\text{CrossEntropy}(P(\mathbf{y}_{ar}^i|\mathbf{x}_{in}^i),\ \mathbf{y}_{ar}^i), 
\end{equation}
where ${\mathbf{y}_{ar}^i}$ is an accent label.

\subsection{Improving ASR with AR}
\label{sec:Accent Embedding Choosing and Transfer}
In a multi-accent ASR-AR, high-quality accent information can guide the ASR task to learn accent-specific pronunciation and expression. Fully utilizing accent information could significantly improve the performance of the ASR task. Therefore, it is worthwhile to study better approaches for utilizing accent information.

In a multi-task ASR-AR, accent embeddings can be chosen from the hidden embedding before the last linear layer ${\mathbf{x}_{dnn}}$, the posterior probabilities of classification ${\mathbf{x}_{pp}}$, or the accent shifts ${\mathbf{s}}$. These three choices can be denoted as:
\begin{equation}
    \label{eq:embedding}
    \left\{\begin{matrix}
    \mathbf{emb}_{dnn}=\mathbf{x}_{dnn}, \hfill \\
    \mathbf{emb}_{pp}=\text{UpProject}(\mathbf{x}_{pp}), \hfill \\
    \mathbf{emb}_{sim}=\text{UpProject}(\mathbf{s}), \hfill
    \end{matrix}\right.
\end{equation}
where ${\text{UpProject}}$ represents a dimension expanding of the embeddings by a linear layer. Since the dimension of ${\mathbf{x}_{dnn}}$ is significantly larger than that of ${\mathbf{x}_{pp}}$ and ${\mathbf{s}}$, for a fair comparison, we use the up-project operation.
The ${\mathbf{emb}_{dnn}}$ provides rich and stable accent information, the ${\mathbf{emb}_{pp}}$ is intuitive and concise, and using the ${\mathbf{emb}_{sim}}$ can provide text-related frame-level accent viriations. In this paper, we compare these three accent embeddings in experiments to determine which one is better.

Moreover, we also investigate the effectiveness of different fusion strategies between accent and the ASR task.
In our triple-encoder scheme, accent information can be fused both implicitly and explicitly. Implicit fusion occurs when the shared encoder learns to extract accent information through the back-propagation of the accent branch, even without the help of an accent embedding. Explicit fusion occurs when an accent embedding is integrated into the attention branch for the ASR task. In the multi-task ASR-AR, implicit fusion is inevitable due to the existence of the accent branch, while explicit fusion is optional. We classify accent fusion into four schemes based on the use of an accent embedding:
\begin{itemize}
    \item ${\mathbf{AF}_{i}}$: Only implicit accent fusion is used in ASR.
    \begin{equation}
        \left\{\begin{matrix}
        \mathbf{x}_{ae}=\text{Encoder}_{att}(\mathbf{x}_{se}), \hfill \\
        P(\mathbf{y}_{c}|\mathbf{x}_{in})=\text{Decoder}_{att}(\mathbf{x}_{ae},\ \mathbf{y}_{c}). \hfill
        \end{matrix}\right. 
    \end{equation}
    \item ${\mathbf{AF}_{ie}}$: Both implicit and explicit (to encoder) accent fusions are used in ASR.
    \begin{equation}
        \left\{\begin{matrix}
        \mathbf{x}_{ae}=\text{Encoder}_{att}(\text{Concat}(\mathbf{x}_{se},\ \mathbf{emb})), \hfill \\
        P(\mathbf{y}_{c}|\mathbf{x}_{in})=\text{Decoder}_{att}(\mathbf{x}_{ae},\ \mathbf{y}_{c}), \hfill
        \end{matrix}\right. 
    \end{equation}
    \item ${\mathbf{AF}_{id}}$: Both implicit and explicit (to decoder) accent fusions are used in ASR.
    \begin{equation}
        \left\{\begin{matrix}
        \mathbf{x}_{ae}=\text{Encoder}_{att}(\mathbf{x}_{se}), \hfill \\ P(\mathbf{y}_{c}|\mathbf{x}_{in})=\text{Decoder}_{att}(\text{Concat}(\mathbf{x}_{ae},\ \mathbf{emb}),\ \mathbf{y}_{c}), \hfill
        \end{matrix}\right. 
    \end{equation}
    \item ${\mathbf{AF}_{ied}}$: Both implicit and explicit (to encoder and decoder) accent fusions are used in ASR.
    \begin{equation}
        \left\{\begin{matrix}
        \mathbf{x}_{ae}=\text{Encoder}_{att}(\text{Concat}(\mathbf{x}_{se},\ \mathbf{emb})), \hfill \\
        P(\mathbf{y}_{c}|\mathbf{x}_{in})=\text{Decoder}_{att}(\text{Concat}(\mathbf{x}_{ae},\ \mathbf{emb}),\ \mathbf{y}_{c}), \hfill
        \end{matrix}\right. 
    \end{equation}
\end{itemize}

In the aforementioned schemes, the effectiveness of ${\mathbf{AF}_{i}}$, ${\mathbf{AF}_{ie}}$, and ${\mathbf{AF}_{id}}$ has been proven in different studies~\cite{zhang2021e2e,deng2021improving,imaizumi2020dialect}. To the best of our knowledge, ${\mathbf{AF}_{ied}}$ has not been extensively explored. However, considering the encoder and decoder respectively focus on linguistic and acoustic information, the ${\mathbf{AF}_{ied}}$ incorporation has the potential to achieve optimal performance. Thanks to the triple-encoder structure, we can easily apply scheme ${\mathbf{AF}_{ied}}$ to the attention branch.

\subsection{Two-granularity Rescoring}
After the decoupling, within our DIMNet, the CTC decoder focuses on pronunciation while the attention decoder emphasizes linguistic information. By leveraging the complementarity between these two types of information, a two-granularity rescoring approach is expected to achieve a better ASR performance compared to single-granularity rescoring.
However, the current mainstream CTC rescoring~\cite{watanabe2017hybrid,li2020comparison} and attention rescoring~\cite{yao2021wenet} necessitate matching sequence lengths between the modeling units of both decoders, making them unsuitable for direct application in two-granularity scenarios. In order to solve this problem, we develop a two-granularity rescoring method based on the CTC rescoring technique~\cite{watanabe2017hybrid} to merge scores from both the CTC and attention decoder.

\begin{figure}[htp]
\centering
\vspace{0em}
\includegraphics[width=20em]{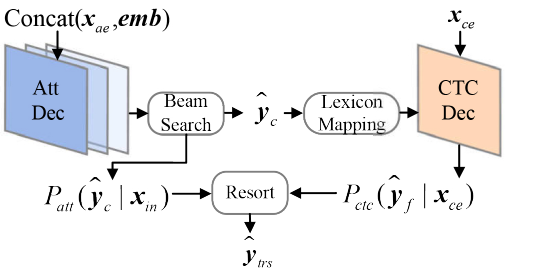}
\vspace{0em}
\caption{Details of two-granularity rescoring.}
\vspace{-1em}
\label{fig:mrs}
\end{figure}

As shown in Fig.~\ref{fig:mrs}, the two-granularity rescoring method involves two-pass decoding. In the first pass, ${N}$ hypotheses ${\hat{\mathbf{y}}_{c}}$ and their corresponding posterior probability scores ${P_{att}(\hat{\mathbf{y}}_{c}|\mathbf{x}_{in})}$ are obtained through autoregressive beam search decoding. The modeling units of these hypotheses are coarse-grained, which can be mapped into fine-grained units through a lexicon, without the need for training a translation model. This process can be represented as:
\begin{equation}
\resizebox{.99\hsize}{!}{$
\left\{\begin{matrix}
    \hat{\mathbf{y}}_{c},\ P_{att}(\hat{\mathbf{y}}_{c}|\mathbf{x}_{in})=\text{BeamSearch}_{att}(\text{Concat}(\mathbf{x}_{ae},\ \mathbf{emb})), \hfill \\
    \hat{\mathbf{y}}^i_{f}=\text{Lexicon}(\hat{\mathbf{y}}^i_{c}), \hfill
\end{matrix}\right.$}
\end{equation}
where ${i\in [1,N]}$ and ${N}$ represents the beam size  used for beam search decoding.

Given the fine-grained hypotheses ${\hat{\mathbf{y}}_{f}}$, we can use the CTC forward algorithm to calculate the scores in the second pass. According to~\cite{graves2006connectionist}, the CTC forward algorithm computes the negative logarithm of the conditional probabilities of the CTC encoder outputs ${\mathbf{x}_{ce}}$ and a given text. If the given text is the transcription label ${\mathbf{y}_{f}}$, the CTC loss is computed. However, if the given text is the hypothesis ${\hat{\mathbf{y}}_{f}}$, the conditional probabilities of ${\mathbf{x}_{ce}}$ and ${\hat{\mathbf{y}}_{f}}$ can be calculated and used for rescoring. Specifically, the conditional probabilities ${\log{P_{ctc}}(\hat{\mathbf{y}}_{f}|\mathbf{x}_{ce})}$ can be computed as:
\begin{equation}
\left\{\begin{matrix}
   \text{CTC}(\mathbf{x}_{ce},\ \hat{\mathbf{y}}_{f})=-\log{P_{ctc}}(\hat{\mathbf{y}}_{f}|\mathbf{x}_{ce}), \hfill \\
   \Downarrow  \\  \log{P_{ctc}}(\hat{\mathbf{y}}_{f}|\mathbf{x}_{ce})=-\text{CTC}(\mathbf{x}_{ce},\ \hat{\mathbf{y}}_{f}). \hfill
\end{matrix}\right. 
\end{equation}

Finally, the ASR prediction ${\hat{\mathbf{y}}_{trs}}$ using the two-granularity rescoring method is obtained by:
\begin{equation}
\begin{split}
    \hat{\mathbf{y}}_{trs}=&\arg\max(w_{1}\log{P_{att}}(\hat{\mathbf{y}}_{c}|\mathbf{x}_{in})\ +\\ &w_{2}\log{P_{ctc}}(\hat{\mathbf{y}}_{f}|\mathbf{x}_{ce})\ +\ w_{3}\log{P_{lm}}(\hat{\mathbf{y}}_{c})),
\end{split}
\end{equation}
where ${w_{1}}$, ${w_{2}}$, and ${w_{3}}$ are tunable parameters. Here, ${P_{lm}(\hat{\mathbf{y}}_{c})}$ is optional language model scores.

\section{Experiments Setup}
\label{sec:Experiments Setup}
\subsection{Dataset}
We conduct extensive experiments on publicly-available English and Chinese datasets to evaluate the proposed DIMNet. For the English experiments, we use the AESRC dataset~\cite{shi2021accented}, while for the Chinese experiments, we use the KeSpeech dataset~\cite{tang2021kespeech}. Details of the two datasets are shown in Table~\ref{tab:datasets}. Besides, following the setup of~\cite{shi2021accented,zhang2021e2e,huang2021aispeech}, the additional Librispeech~\cite{panayotov2015librispeech} dataset is also used for the AESRC experiments.

\begin{table}[h]
    \centering
    \footnotesize
    \caption{Details of the English AESRC and the Chinese KeSpeech accent datasets.}
    \label{tab:datasets}
    \begin{tabular}{@{}ccccc@{}}
    \toprule
    \textbf{Dataset} & \textbf{\begin{tabular}[c]{@{}c@{}}Accent \\ Num\end{tabular}} & \textbf{\begin{tabular}[c]{@{}c@{}}Total Duration\\ (Hours)\end{tabular}} & \textbf{\begin{tabular}[c]{@{}c@{}}Sampling Rate\\ (kHz)\end{tabular}} & \textbf{Style} \\ \midrule
    AESRC            & 8                                                              & 160                                                                       & 16                                                                 & Reading        \\ \midrule
    KeSpeech         & 9                                                              & 1542    & 16                                                                 & Reading        \\ \bottomrule
    \end{tabular}
\end{table}

In this study, we use BPE-phoneme as the two-granularity modeling units for English, and char-syllable for Chinese. This choice is based on the unique characteristics of each language and the effectiveness of these units in capturing phonetic and semantic features. To convert the coarse-grained units to fine-grained
units, we utilize CMUdict\footnote{Available at \url{http://www.speech.cs.cmu.edu/cgi-bin/cmudict}} and Pypinyin\footnote{Available at \url{https://pypi.org/project/pypinyin/}} lexicons, respectively. Refer to Table~\ref{tab:two_level_units} for detailed information.
\begin{table}[h]
    \centering
    \footnotesize
    \caption{Two-granularity modeling units details.}
    \label{tab:two_level_units}
    \begin{tabular}{@{}cccc@{}}
    \toprule
    \multirow{2}{*}{\textbf{Language}} & \multirow{2}{*}{\textbf{\begin{tabular}[c]{@{}c@{}}Coarse-grained \\ Units Num\end{tabular}}} & \multirow{2}{*}{\textbf{\begin{tabular}[c]{@{}c@{}}Fine-grained \\ Units Num\end{tabular}}} & \multirow{2}{*}{\textbf{Lexicon}} \\
    &  &  &  \\ \midrule
    English & 5002 (BPE) & 40 (Phoneme) & ${\text{CMUdict}^1}$ \\ \midrule
    Chinese & 5687 (Char) & 419 (Syllable) & ${\text{Pypinyin}^2}$ \\ \bottomrule
    \end{tabular}
\end{table}

\subsection{Model Configurations}
Our baseline model is based on a triple-branch structure, which utilizes the hidden embedding before the last linear layer in the accent branch to generate accent embeddings. The encoders are based on Conformer~\cite{gulati2020conformer}, while the attention decoder is Transformer~\cite{vaswani2017attention}. The shared encoder, CTC encoder, and attention encoder are comprised of 9, 3, and 3 Conformer blocks, respectively, while the attention decoder consists of 6 blocks. 
The Conformer blocks have 2048 inner dimensions for feed-forward networks (FFN), 256 model dimensions, 4 attention heads, and utilize a CNN kernel size of 15.
During training, the loss weights assigned to the CTC, AR, and ASR branches are 0.3, 0.4, and 0.3, respectively. We specifically lightly amplify the AR loss weight due to its relatively smaller absolute loss value.
In addition, for the LASAS AR model, we set the mapping spaces to 8, and the rest parameters are the same as~\cite{shao2022linguistic}. Our experiments include SpecAugment, model average, and a 2-layer-Transformer based language model (LM). 

In English experiments, our model is first trained for 70 epochs on the mixture of the AESRC and LibriSpeech datasets and then fine-tuned 50 epochs on the AESRC alone. In the first stage, we update the CTC and attention branches using both the AESRC and LibriSpeech datasets, while only the AESRC dataset is used to update the accent branch. In the Chinese experiments, we train the model for 100 epochs without finetuning. Similarly, we use all the available data to update the CTC and attention branches, while only the accent data is used to update the accent branch. It is worth noting that the accent data referred to here is denoted as Phase 1 in~\cite{tang2021kespeech}, which includes a subset of Mandarin. To account for the large differences in data amount for different accents in the KeSpeech dataset, we use unbalanced weights for the CE loss in the AR task, which is set as the ratio of the number of accent utterances with respect to the Mandarin subset.

\section{Experiments Results}
\label{sec:Experiments Results}
This section presents the experimental results of our proposed DIMNet. We first introduce the results of baselines and our ablation and comparison experiments on the AESRC dataset to demonstrate the effectiveness of each module. Next, we compare our approach to the general A2P+P2W cascade modeling schemes of two-granularity units. Finally, we compare the performance of DIMNet to previous studies on both the AESRC and KeSpeech datasets.

\subsection{Effectiveness of DIMNet}
In Table~\ref{tab:ablation_and_contrast}, ${B1-B4}$ represent CTC/attention ASR and multi-task ASR-AR baselines, while ${D1}$ represents our proposed DIMNet. 
Specifically, ${B1}$ is a classic CTC/attention ASR model using only coarse-grained BPE units, and ${B2}$ is derived from ${B1}$ by replacing the BPE units with phonemes in the CTC branch. Both ${B1}$ and ${B2}$ have no AR branch. 
${B3}$ is a widely-used multi-task ASR-AR model as described in study~\cite{zhang2021e2e}, which comprises a shared encoder and three branches: CTC decoder, AR, and attention decoder. The AR branch incorporates pooling and linear layers. In addition, to ensure a fair comparison, we incorporate the accent embedding fusion into the attention decoder with the detach operation.
In ${B4}$, we substitute the shared encoder of ${B3}$ with the triple encoder structure utilized in DIMNet. Essentially, ${B4}$ can be viewed as replacing the LASAS AR of DIMNet with a basic AR model.

The comparison between ${B1}$ and ${B2}$ highlights the effect of two-granularity units. The direct introduction of phonemes in a CTC/attention framework can only lead to a slight improvement in ASR performance. After integrating an AR branch, the ASR performance of ${B3}$ enhanced compared to ${B2}$, which is a widely verified phenomenon. Comparing ${B4}$ with ${B3}$, the introduction of the triple encoder structure yields a mere average relative impact of ${0.15\%}$ on accent accuracy, while it improves the ASR task by an average relative ${1.02\%}$. Ultimately, when comparing ${D1}$ with ${B1-B4}$, it becomes evident that DIMNet exhibits clear advantages in both AR and ASR, underscoring the benefits of DIMNet beyond modeling units and multi-task learning.

\begin{table*}[]
\centering
\caption{Ablation and contrast experiments of our proposed DIMNet on the AESRC dataset. See Section\ref{sec:Accent Embedding Choosing and Transfer} for the definitions of accent embedding (${\mathbf{emb}}$) and accent fusion (${\mathbf{AF}}$). In the third column of the table, G, P, and B represent grapheme, phoneme, and BPE, respectively.}
\label{tab:ablation_and_contrast}
\begin{tabular}{@{}cccccccccccc@{}}
\toprule
\multirow{2}{*}{\textbf{ID}} & \multirow{2}{*}{\textbf{Model}} & \multirow{2}{*}{\textbf{\begin{tabular}[c]{@{}c@{}}CTC/ATT\\ Units\end{tabular}}} & \multirow{2}{*}{\textbf{\begin{tabular}[c]{@{}c@{}}Triple\\ Encoders\end{tabular}}} & \multirow{2}{*}{\textbf{LASAS}} & \multirow{2}{*}{\textbf{\begin{tabular}[c]{@{}c@{}}Accent\\ Embedding\end{tabular}}} & \multirow{2}{*}{\textbf{\begin{tabular}[c]{@{}c@{}}Accent\\ Fusion\end{tabular}}} & \multirow{2}{*}{\textbf{Rescoring}} & 
\multicolumn{2}{c}{\textbf{\begin{tabular}[c]{@{}c@{}}AR\\ ACC (\%)\end{tabular}}} & \multicolumn{2}{c}{\textbf{\begin{tabular}[c]{@{}c@{}}ASR\\ WER (\%)\end{tabular}}} \\ \cmidrule(l){9-12} 
& & & & & & & & \textbf{Dev} & \textbf{Test} & \textbf{Dev} & \textbf{Test} \\ \hline\hline
B1 &  & B/B & \xmark &  & \xmark & \xmark &  & - & - & 5.96 & 6.91 \\
B2 &  & P/B & \xmark &  & \xmark & \xmark &  & - & - & 5.96 & 6.88 \\
B3 &  & P/B & \xmark &  & ${\mathbf{emb}_{dnn}}$ & ${\mathbf{AF}_{id}}$ &  & 79.12 & 74.02 & 5.83 & 6.76 \\
B4 & \multirow{-4}{*}{\begin{tabular}[c]{@{}c@{}}CTC/ATT \\ AR-ASR\\ Baseline\end{tabular}} & P/B & \cmark & \multirow{-4}{*}{\xmark} & ${\mathbf{emb}_{dnn}}$ & ${\mathbf{AF}_{id}}$ & \multirow{-4}{*}{\xmark} & 79.08 & 73.84 & 5.78 & 6.68 \\ \midrule
D1 & \begin{tabular}[c]{@{}c@{}}DIMNet\\ Basline\end{tabular} & P/B & \cmark & \cmark & ${\mathbf{emb}_{dnn}}$ & ${\mathbf{AF}_{ied}}$ & \xmark & 86.47 & 78.82 & 5.55 & 6.27 \\ \midrule
D2 & \multirow{3}{*}{\begin{tabular}[c]{@{}c@{}}Decoupling\\ Ablation\end{tabular}} & G/B & \cmark & \multirow{3}{*}{\cmark} & \multirow{3}{*}{${\mathbf{emb}_{dnn}}$} & \multirow{3}{*}{${\mathbf{AF}_{ied}}$} & \multirow{3}{*}{\xmark} & 81.52 & 73.58 & 5.54 & 6.38 \\
D3 & & B/B & \cmark &  &  &  &  & 81.9 & 75.68 & 5.74 & 6.41 \\
D4 & & P/B & \xmark & & & & & 86.15 & 79.78 & 5.87 & 6.63 \\ \midrule
D5 &  &  &  & w/o Text Input &  &  &  & 76.54 & 71.28 & 5.88 & 6.61 \\
D6 &  &  &  & w/o Detach &  &  &  & 80.47 & 74.67 & 5.55 & 6.32 \\
D7 & \multirow{-3}{*}{\begin{tabular}[c]{@{}c@{}}AR\\ Ablation\end{tabular}} & \multirow{-3}{*}{P/B} & \multirow{-3}{*}{\cmark} & w/o Frame Level & \multirow{-3}{*}{${\mathbf{emb}_{dnn}}$} & \multirow{-3}{*}{${\mathbf{AF}_{ied}}$} & \multirow{-3}{*}{\xmark} & 85.7 & 80.15 & 5.55 & 6.3 \\ \midrule
D8 &  &  &  &  & \xmark & ${\mathbf{AF}_{i}}$ &  & 85.04 & 80.15 & 5.71 & 6.49 \\
D9 &  &  &  &  & ${\mathbf{emb}_{pp}}$ & ${\mathbf{AF}_{ied}}$ &  & 85.25 & 79.4 & 5.52 & 6.32 \\
D10 &  &  &  &  & ${\mathbf{emb}_{sim}}$ & ${\mathbf{AF}_{ied}}$ & & 84.88 & 79.42 & 5.55 & 6.3 \\
D11 &  &  &  &  & ${\mathbf{emb}_{dnn}}$ & ${\mathbf{AF}_{ie}}$ &  & 85.11 & 79.4 & 5.47 & 6.36 \\
D12 & \multirow{-5}{*}{\begin{tabular}[c]{@{}c@{}}ASR\\ Ablation\end{tabular}} & \multirow{-5}{*}{P/B} & \multirow{-5}{*}{\cmark} & \multirow{-5}{*}{\cmark} & ${\mathbf{emb}_{dnn}}$ & ${\mathbf{AF}_{id}}$ & \multirow{-5}{*}{\xmark} & 85.89 & \textbf{80.2} & 5.64 & 6.4 \\ \midrule
D13 & \begin{tabular}[c]{@{}c@{}}Two-Gran\\ Rescoring\end{tabular} & P/B & \cmark & \cmark & ${\mathbf{emb}_{dnn}}$ & ${\mathbf{AF}_{ied}}$ & \cmark & \textbf{86.47} & 78.82 & \textbf{5.41} & \textbf{6.13} \\ \bottomrule
\end{tabular}
\end{table*}
 
\subsection{Decoupling of AR and ASR}
In ${D2}$ and ${D3}$, we examine the influence of modeling units on DIMNet. Specifically, we employ grapheme-BPE instead of phoneme-BPE units in ${D2}$, while utilizing BPE-BPE units in ${D3}$. Comparing ${D1}$ and ${D2}$, it becomes apparent that not all fine-grained units are compatible with DIMNet. Graphemes, in comparison to phonemes, exhibit semantic relevance but lack adequate pronunciation information. As a result, they are unable to effectively enhance the performance of AR tasks, consequently indirectly diminishing ASR performance.
When comparing ${D1}$ and ${D3}$, it is clear that utilizing phonemes-BPE two-granularity units leads to enhanced performance for DIMNet in terms of both accent accuracy (ACC) and word error rate (WER). However, if coarse-grained BPE units are employed in both the CTC and attention branches, we observe a significant decrease in results. Specifically, there is a relative drop of ${5.29\%}$ and ${3.98\%}$ in the AR task, as well as a relative drop of ${3.42\%}$ and ${2.23\%}$ in the ASR task.
This result clearly demonstrates that decoupling the modeling units is a crucial factor in improving both AR and ASR performance, even when utilizing the same interactive structure.
However, the use of a two-granularity modeling unit presents challenges for the ASR task without the triple-encoder structure. 
To demonstrate this, we consider the setup of ${D4}$, where we remove the CTC encoder and attention encoder in Fig.~\ref{fig:system_framework}, keeping only a 12-layer shared encoder, and directly put the accent embedding to the attention decoder. 
Comparing ${D4}$ and ${D1}$, we observe that while ${D4}$ achieves a similar AR accuracy to ${D1}$, its ASR performance significantly degrades by ${5.77\%}$ and ${5.74\%}$ respectively. 
This finding suggests that the triple-encoder structure effectively decouples the modeling process of different units, enabling the CTC and attention branches to output units with different granularity.

\begin{figure}[h]
\centering
\vspace{0em}
\includegraphics[width=24em]{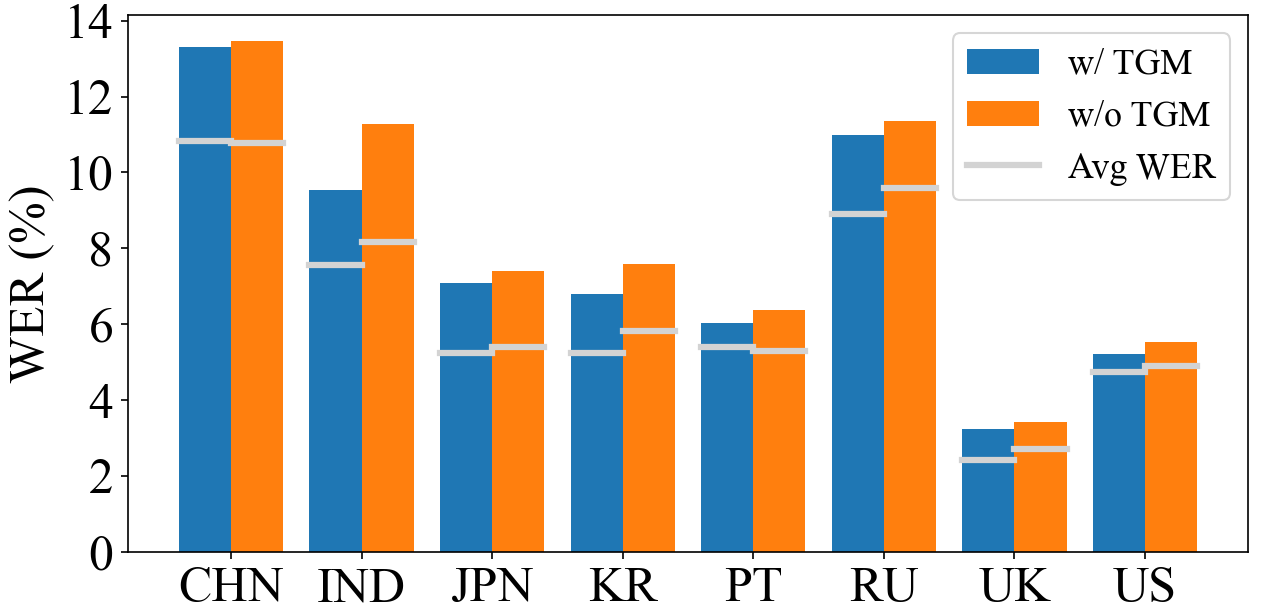}
\vspace{0em}
\caption{WER for the word which contains the top 5 PER phonemes.}
\vspace{-1em}
\label{fig:Top5_WER}
\end{figure}

Generally, accent utterances not only contain accent words but also include standard pronounced common words. Therefore, we conduct further analysis to investigate whether performance improvement of the ASR and AR is evident in accent words.
Table~\ref{tab:top5_per} illustrates that the use of fine-grained phoneme units can improve the accuracy of all accents in the AR task. This suggests that fine-grained phonemes are more effective for AR tasks than coarse-grained BPE, and this advantage applies to all accents. 
We assume that the phonemes with the highest PER in the CTC branch represent accent pronunciation, and list them in Table~\ref{tab:top5_per}. As shown in the table, these high-PER phonemes align with our general knowledge and previous findings in accent linguistic research~\cite{yarra2019indic,igarashi2020improving,han2013pronunciation}. This finding suggests that the phonemes which cause difficulties for the DIMNet indeed contain accents, and that our proposed model effectively captures accent-specific pronunciation-related information. 
Furthermore, we count the WERs of the attention branch for words that contain the top 5 PER phonemes which are listed in Table~\ref{tab:top5_per} and compare them to the average WERs of each accent in Fig.~\ref{fig:Top5_WER}. 
As depicted in Fig.~\ref{fig:Top5_WER}, the WERs of these accent words are higher than the average WERs, suggesting that they are more difficult to recognize in the ASR task, and hence contribute to the increase in the average WER. However, the introduction of two-granularity units results in a decrease in the WER of difficult words in each accent, compared to the case without it. This finding clearly demonstrates the effectiveness of our proposed scheme in improving the recognition of difficult words in different accents.
\begin{table}[ht]
\centering
\footnotesize
\caption{Impact of fine-grained units on AR accuracy, as well as the fine-grained units with the highest PER. TGM in the table refers to two-granularity modeling. Only the results of the AESRC test set are shown.}
\label{tab:top5_per}
\begin{threeparttable}
\begin{tabular}{@{}cccc@{}}
\toprule
\multirow{2}{*}{\textbf{Accent}} & \multicolumn{2}{c}{\textbf{AR ACC (\%)}} & \multirow{2}{*}{\textbf{Phonemes of Top 5 PER}\tnote{\S}} \\ \cmidrule(lr){2-3}
 & \textbf{w/ TGM\tnote{\dag}} & \textbf{w/o TGM\tnote{\ddag}} &  \\ \hline\hline
CHN & 80.68 & 79.98 & [ZH], [EH], [AO], [TH], [EY] \\ \midrule
IND & 93.30 & 91.05 & [OY], [ZH], [SH], [JH], [TH] \\ \midrule
JPN & 72.30 & 68.90 & [ZH], [L], [OW], [R], [AO] \\ \midrule
KR & 83.15 & 79.39 & [ZH], [OW], [AE], [AO], [UH] \\ \midrule
PT & 80.76 & 75.92 & [ZH], [UH], [AE], [AO], [EH] \\ \midrule
RU & 74.49 & 70.39 & [OY], [AE], [JH], [OW], [UH] \\ \midrule
UK & 94.32 & 93.49 & [NG], [AO], [OW], [ER], [AA] \\ \midrule
US & 58.36 & 53.64 & [UH], [AE], [AA], [ZH], [OW] \\ \bottomrule
\end{tabular}
\begin{tablenotes}
    \centering
    \footnotesize
    \item{\dag}: This model is ${D1}$ in the Table~\ref{tab:ablation_and_contrast}.
    \item{\ddag}: This model is ${D3}$ in the Table~\ref{tab:ablation_and_contrast}. 
    \item{\S}: Correspondence between ARPABET and IPA phoneme sets: \url{https://en.wikipedia.org/wiki/ARPABET}
\end{tablenotes}
\end{threeparttable}
\end{table}

\subsection{Improving AR with ASR}
\label{sec:ASR’s assistance to AR}
To investigate the impact of ASR on AR, we conduct the ${D5}$ experiment presented in Table~\ref{tab:ablation_and_contrast}. In this experiment, we remove the text input of the LASAS-based accent branch in the ${D1}$ model and replace it with the outputs of the shared encoder. This design allowed us to eliminate the fusion of linguistic information from the ASR to the AR while keeping the total parameters of the DIMNet model unchanged.
After comparing the performance of ${D1}$ and ${D5}$, it is apparent that the exclusion of linguistic information from the CTC branch has a substantial impact on the accuracy of the accent branch, resulting in a decrease in the AR performance. This outcome clearly demonstrates the significance of the linguistic information from the CTC branch in enhancing the performance of AR tasks.

Furthermore, we evaluate the modifications made to the original LASAS AR model~\cite{shao2022linguistic} for its adaptation to the multi-task framework. Comparison of ${D6}$ and ${D1}$ reveals that the detach operation of linguistic inputs and accent outputs in Fig.~\ref{fig:system_framework} of the accent branch has a significant positive impact on the performance of AR. In both the dev and test sets, the relative improvements are up to ${6.94\%}$ and ${5.27\%}$, respectively. The detach operation enables the AR task to focus solely on accent-specific information, thereby improving the effectiveness of accent branch optimization. Additionally, the detach operation reduces interference from the AR task to the ASR task, thereby improving ASR performance to some extent. These results demonstrate the significance of the detach operation of the accent branch in multi-task ASR-AR. 
By comparing ${D7}$ and ${D1}$, we can analyze the impact of frame-level and utterance-level CE loss on accent prediction. The AR accuracy of ${D7}$ is slightly lower than that of ${D1}$ in the dev set, but the opposite is observed in the test set. This suggests that the utterance-level loss used in ${D7}$ can enhance the model's generalization and mitigate overfitting. However, the frame-level accent information used in ${D1}$ achieves a slightly better performance in the ASR task. We believe this is because the frame-level accent information can help correct fine-grained errors caused by accents in the ASR task. In our experience, accent speech utterances often consist of a small portion of words that exhibit accent characteristics, while the remaining words are pronounced in a standard manner. In other words, the accent words in a sentence pose a challenge for the ASR task but are easier to recognize for the AR task. Therefore, although utterance-level pooling operation indeed improves the overall accuracy in the AR task, it may not capture the distinguishing details of the accent words as effectively as without pooling. Hence, the selection of frame-level or utterance-level CE loss depends on the practical application and trade-off between the AR and ASR tasks.

\subsection{Improving ASR with AR}
We further investigate the benefits of incorporating AR information into the ASR task, similar to Section~\ref{sec:ASR’s assistance to AR}. We first consider the ${D8}$ experiment in Table~\ref{tab:ablation_and_contrast}, where no accent embedding is fused to the attention branch. Comparing ${B1}$ and ${D8}$, we find that although ${D8}$ lacks explicit accent embedding, it can still implicitly extract accent information through the shared encoder. This leads to significant improvements in the ASR task compared to ${B1}$. Specifically, the relative improvement in implicit accent fusion on the ASR dev and test sets is approximate ${4.19\%}$ and ${6.08\%}$, respectively. However, when comparing ${D8}$ with ${D1}$, we observe that explicit fusion of the accent embedding to the ASR branch leads to better results, which underscores the value of AR-to-ASR interaction.

Next, the impact of different accent embeddings is analyzed. ASR performance can be used to evaluate the quality of accent embedding since it serves the ASR task. The results in Table~\ref{tab:ablation_and_contrast} show that ${\mathbf{emb}_{dnn}}$ performs the best, while ${\mathbf{emb}_{sim}}$ performs moderately well, and ${\mathbf{emb}_{pp}}$ performs the worst. ${\mathbf{emb}_{dnn}}$ contains utterance-level accent classification information that is relatively stable, which is easier for an ASR model to recognize. On the other hand, ${\mathbf{emb}_{pp}}$ loses a significant amount of acoustic information valuable to the ASR model due to compression by the final DNN layer. In addition, although ${\mathbf{emb}_{sim}}$ and ${\mathbf{emb}_{pp}}$ have the same dimensions in our experiments, ${\mathbf{emb}_{sim}}$ is better. 
This finding suggests that accent shifts ${\mathbf{emb}_{sim}}$ also have utility in improving ASR performance. However, since ${\mathbf{emb}_{sim}}$ operate at the frame level, they are inherently more complex and variable, which presents a challenge for ASR models to effectively leverage this information.

In experiments ${D11}$ and ${D12}$, we compare two additional accent fusion schemes ${\mathbf{AF}_{ie}}$ and ${\mathbf{AF}_{id}}$, both of which explicitly integrate an accent embedding into the attention branch of the ASR task. Comparing the performance of ${D1}$, ${D11}$, and ${D12}$, we find that using an encoder to process an accent embedding is more effective than using a decoder, but the best performance is achieved when both are used. The effectiveness of using an encoder to fuse accent embeddings is widely acknowledged, as the encoder can focus on acoustic information and be adjusted based on accent embedding. This enables the generation of a context representation that is easier for the decoder to understand.
Although there are limited studies on fusing accent embeddings to a decoder, our experiments suggest that it is also an effective approach, as the decoder focuses more on linguistic information, and accent embeddings can help correct accent-specific words. The complementary roles of the encoder and decoder in utilizing accent embeddings explain why ${\mathbf{AF}_{ied}}$ outperforms both ${\mathbf{AF}_{ie}}$ and ${\mathbf{AF}_{id}}$. Therefore, integrating accent embeddings into both the encoder and decoder is essential to achieve better performance. Thanks to the triple-encoder structure, this can be easily accomplished in the attention branch.

\subsection{Two-granularity Rescoring}
In Table~\ref{tab:Tab_TRS_RTF}, we evaluate the effectiveness of two-granularity rescoring. As shown in the table, attention rescoring yields an average relative reduction in WER of ${1.63\%}$ within the CTC/attention framework. To mitigate the framework's impact, we also assessed the effects of CTC rescoring on the DIMNet with single-granularity BPE units. The average WER decrease observed in this case is similar to that of attention rescoring, approximately relative ${1.41\%}$. In comparison to these two classic rescoring techniques, our proposed two-granularity rescoring achieves an average relative WER decrease of ${2.38\%}$. In the DIMNet model, the CTC and attention branches operate independently, and each branch has a different and complementary focus in terms of information granularity. Two-granularity rescoring effectively integrates the prediction scores of the two-granularity units, thereby complementing the limitations of the decoupling operation and leading to further improvements in the DIMNet model's performance.
\begin{table}[]
    \centering
    \footnotesize
    \caption{Effectiveness of two-granularity rescoring. In this table, ARS, CRS and TRS represent attention rescoring, CTC rescoring, and two-granularity rescoring decoding, respectively.}
    \label{tab:Tab_TRS_RTF}
    \begin{threeparttable}
    \begin{tabular}{@{}llccccc@{}}
    \toprule
    \multirow{2}{*}{\textbf{ID}} & \multirow{2}{*}{\textbf{Model}} & \multicolumn{2}{c}{\textbf{Decode Pass}} & \multicolumn{2}{c}{\textbf{WER (\%)}} \\ \cmidrule(l){3-6} 
    &  & \textbf{1-st} & \textbf{2-nd} & \textbf{Dev} & \textbf{Test} \\ \midrule
    B1 & \multirow{2}{*}{CTC/ATT} & ATT & - & 5.96 & 6.91 \\
    B1+ARS &  & CTC & ATT & 5.86 & 6.80 \\ \midrule
    D3 & \multirow{2}{*}{DIMNet w/o TGM} & ATT & - & 5.74 & 6.41 \\
    D3+CRS &  & ATT & CTC & 5.65 & 6.33 \\ \midrule
    D1 & \multirow{2}{*}{DIMNet} & ATT & - & 5.55 & 6.27 \\
    D1+TRS\tnote{\dag} &  & ATT & CTC & 5.41 & 6.13 \\ \bottomrule
    \end{tabular}
    \begin{tablenotes}
        \centering
        \footnotesize
        \item{\dag}: This model is ${D13}$ in the Table~\ref{tab:ablation_and_contrast}. 
    \end{tablenotes}
    \end{threeparttable}
\end{table}

\subsection{A2P+P2W vs. A2P+A2W for Two-granularity Units}
As mentioned in Section \ref{sec:Related Works}, a common approach for two-granularity modeling in ASR involves phoneme recognition and translation of phonemes into words or BPEs (A2P+P2W). However, this approach faces the challenges of error accumulation in the P2W stage. In contrast, the DIMNet directly recognizes phonemes and BPEs from audio features (A2P+A2W). In this section, we compare the two approaches.
\begin{table}[ht]
    \centering
    \footnotesize
    \caption{Comparison of A2P+P2W and A2P+A2W approaches to modeling two-granularity units. Only the AESRC test set results are shown.}
    \label{tab:Tab3_translate}
    \begin{threeparttable}
    \begin{tabular}{@{}ccc@{}}
    \toprule
    \textbf{Model} & \textbf{\begin{tabular}[c]{@{}c@{}}CTC Dec\\ PER (\%)\end{tabular}} & \textbf{\begin{tabular}[c]{@{}c@{}}Att Dec\\ WER (\%)\end{tabular}} \\ \hline\hline
    A2P + A2W\tnote{\dag} & 4.41 & 6.27 \\ \midrule
    \begin{tabular}[c]{@{}c@{}}A2P + Soft P2W \end{tabular} & 4.59 & 6.4 \\ \midrule
    \begin{tabular}[c]{@{}c@{}}A2P + Hard P2W\end{tabular} & 4.61 & 7.54 \\ \bottomrule
    \end{tabular}
    \begin{tablenotes}
        \centering
        \footnotesize
        \item{\dag}: This model is ${D1}$ in the Table~\ref{tab:ablation_and_contrast}. 
    \end{tablenotes}
    \end{threeparttable}
\end{table}

Table \ref{tab:Tab3_translate} presents the results of two models that translate soft embeddings and hard OneHot vectors of phonemes into BPE, both based on the DIMNet. In both models, we maintain the triple-encoder structure, but modify the input of the attention encoder to CTC phoneme information instead of the shared encoder outputs, while keeping the accent embedding concatenation unchanged. We aim for the attention branch to act as a P2W model in both schemes. To achieve this, we detach the outputs of the CTC branch as well as the accent branch to ensure that the attention branch solely focuses on translating phonemes into BPE. The results show that the change in PER for the CTC decoder is minimal because detaching the CTC phonemes makes the CTC branch relatively independent, ensuring a fair comparison of the P2W process. As shown in Table \ref{tab:Tab3_translate}, Firstly, the WER of A2P+Soft P2W and A2P+Hard P2W is inferior to that of A2P+A2W, indicating the superiority of the DIMNet. Secondly, when the input phoneme sequence's PER is equivalent, the WER of using soft embeddings is relative ${17.81\%}$ higher than that of using OneHot vectors. This is because the CTC encoder outputs contain richer linguistic and acoustic information, while regular phonemes only contain linguistic information. However, in most two-granularity unit ASR, hard OneHot vectors of phonemes are used as inputs, which limits the performance of P2W. Moreover, the use of completely correct phonemes during training and hypothesis phonemes during inference in P2W can cause a mismatch and error accumulation. In contrast, our triple-encoder scheme independently models phonemes and BPEs by the CTC encoder and attention encoder, respectively, which can help mitigate error accumulation.

\subsection{Comparison with Previous Studies}
\label{sec:Comparison with Published Studies}
In Table~\ref{tab:Tab5_overall_AESRC}, we present a comparison of our DIMNet with several other typical approaches on the AESRC dataset. The second row shows an ASR-AR cascade scheme~\cite{qian2022layer} that achieves top-level performance on this dataset. Comparing it with our DIMNet, we can see that their model achieves a better result in the AR task, which is mainly due to extensive data augmentation. 
Without the data augmentation, their AR accuracy on the dev set is ${84.51\%}$, which is slightly lower than that of the DIMNet. Apart from the second row, the DIMNet significantly outperforms other schemes in the AR task. In particular, the DIMNet surpasses our previous LASAS AR model~\cite{shao2022linguistic}, which demonstrates the value of our improvements in the original LASAS. These results indicate that the DIMNet is highly competitive in AR tasks. In the ASR task, the first row is a CTC/attention ASR, while the second to fourth rows are typical multi-task ASR-AR introduced in Section~\ref{sec:Related Works}. Except for the first row, the rest of the models do not use language models. In a comparable situation, our DIMNet's ASR performance surpasses the above schemes. This indicates that the DIMNet also has significant advantages in the ASR task. Moreover, the last row shows that after adding the LM, the performance of the DIMNet can be further improved. By comparing the DIMNet in the last row and the CTC/attention-based baseline in the first row, we obtain relative improvements of ${21.45\%}$ and ${32.33\%}$ on the test sets of AR and ASR tasks, respectively. This fully demonstrates that our scheme is effective in English.
\begin{table}[htp]
    \centering
    \footnotesize
    \caption{Comparison of different approaches on the AESRC dataset.}
    \label{tab:Tab5_overall_AESRC}
    \begin{threeparttable}
    \begin{tabular}{@{}ccccc@{}}
        \toprule
        \multirow{2}{*}{\textbf{Model}} & \multicolumn{2}{c}{\textbf{\begin{tabular}[c]{@{}c@{}}AR Task\\ ACC (\%)\end{tabular}}} & \multicolumn{2}{c}{\textbf{\begin{tabular}[c]{@{}c@{}}ASR Task\\ WER (\%)\end{tabular}}} \\ \cmidrule(l){2-5} 
         & \textbf{Dev} & \textbf{Test} & \textbf{Dev} & \textbf{Test} \\ \hline\hline
        AESRC Baseline~\cite{shi2021accented} & 76.1 & 64.9 & 6.92 & 8.29 \\ \midrule
        ASR-AR Cascade~\cite{qian2022layer} & \textbf{91.13} & \textbf{83.63} & 5.53 & 6.56 \\ \midrule
        STJR~\cite{zhang2021e2e}: ASR-AR Single-task & 77 & 72.2 & 5.8 & 6.6 \\ \midrule
        MTJR~\cite{zhang2021e2e}: ASR-AR Multi-task & 82.4 & 75.2 & 6.2 & 7.1 \\ \midrule
        LASAS~\cite{shao2022linguistic}: Only AR task & 84.88 & 77.42 & - & - \\ \midrule
        DIMNet w/ TRS\tnote{\dag} & 86.47 & 78.82 & 5.41 & 6.13 \\ \midrule
        DIMNet w/ TRS+LM & 86.47 & 78.82 & \textbf{5.03} & \textbf{5.61} \\ \bottomrule
    \end{tabular}
    \begin{tablenotes}
        \centering
        \footnotesize
        \item{\dag}: This model is ${D13}$ in the Table~\ref{tab:ablation_and_contrast}. 
    \end{tablenotes}
    \end{threeparttable}
\end{table}

Table~\ref{tab:Tab6_overall_KeSpeech} presents the experimental results on the KeSpeech dataset~\cite{tang2021kespeech}. 
In the ASR task, the first row represents an official baseline model~\cite{tang2021kespeech} trained using the Espnet~\cite{watanabe2017hybrid} tool, incorporating an LM. The second and third rows correspond to baselines that we trained ourselves using the Wenet~\cite{yao2021wenet} tools. All three rows are to CTC/attention frameworks.
For the AR task, the KeSpeech baseline~\cite{tang2021kespeech} is a ResNet34~\cite{he2016deep} model. We train the CTC/attention and DIMNet models under a comparable conditions. As shown in the table, the DIMNet outperforms both baselines significantly on both AR and ASR tasks. Specifically, the DIMNet achieves a relative improvement of ${28.53\%}$ over the KeSpeech baseline on the AR task and a relative improvement ${14.55\%}$ on the ASR task, demonstrating the effectiveness of the DIMNet in Chinese. 
Notably, unlike phoneme/BPE units, syllable/char units have the same time steps, indicating the robustness of the DIMNet to the time steps of coarse and fine-grained units, which makes it applicable to other languages as well.
\begin{table}[htp]
    \centering
    \footnotesize
    \caption{Comparison of different approaches on the KeSpeech dataset. The ASR CER in the first line is obtained by the utterance number weighted average of the results in~\cite{tang2021kespeech}. And the AR ACC is the average of 6 accents.}
    \label{tab:Tab6_overall_KeSpeech}
    \begin{tabular}{@{}ccccc@{}}
    \toprule
    \multirow{2}{*}{\textbf{Model}} & \multicolumn{2}{c}{\textbf{\begin{tabular}[c]{@{}c@{}}AR Task\\ ACC (\%)\end{tabular}}} & \multicolumn{2}{c}{\textbf{\begin{tabular}[c]{@{}c@{}}ASR Task\\ CER (\%)\end{tabular}}} \\ \cmidrule(l){2-5} 
     & \textbf{Dev} & \textbf{Test} & \textbf{Dev} & \textbf{Test} \\ \hline\hline
    KeSpeech Baseline~\cite{tang2021kespeech} & - & 61.13 & - & 10.38 \\ \midrule
    CTC/ATT w/ ARS & - & - & 6.05 & 9.54 \\ \midrule
    CTC/ATT w/ ARS + LM& - & - & 5.95 & 9.39 \\ \midrule
    DIMNet w/ TRS & 80.06 & 78.57 & 5.90 & 9.40 \\ \midrule
    DIMNet w/ TRS + LM & \textbf{80.06} & \textbf{78.57} & \textbf{5.71} & \textbf{8.87} \\ \bottomrule
    \end{tabular}
\end{table}

\section{Conclusions}
\label{sec:Conclusions}
In this paper, we propose the DIMNet, a multi-task framework for joint ASR-AR tasks. Our approach first decouples the AR and ASR tasks using a triple-encoder structure that can model two-granularity units in each task. Then we enhance the interaction between the two tasks by introducing and improving the LASAS AR model and studying the selection and fusion of accent embeddings. Finally, we develop a two-granularity rescoring scheme that effectively combines two-granularity scores to further enhance ASR performance. Experimental results demonstrate that our scheme achieves relative improvements in AR accuracy of ${21.45\%}$ and ${28.53\%}$, as well as relative reductions in ASR error rate of ${32.33\%}$ and ${14.55\%}$ on test sets of the AESRC and KeSpeech datasets, respectively, compared to the E2E baselines. Looking forward, we aim to further reduce the computational complexity of the DIMNet and extend its application to multilingual ASR tasks.

\bibliographystyle{Transactions-Bibliography/IEEEtran}
\bibliography{Ref.bib}
%




\end{document}